\documentclass[11pt,a4paper]{article}

\makeatletter 

\usepackage{color,amsmath,amssymb,wrapfig,url,slashed,cite,ifpdf}

\renewcommand\citepunct{,\penalty\@M\hskip.13emplus.1emminus.1em\relax} 

\ifpdf                                     
  \usepackage{hyperref,graphicx}           
\else                                      
  \usepackage[dvipdfmx]{hyperref,graphicx} 
\fi

\newcommand{\dd}{{\mathrm d}}      

\newcommand{\vc}[1]{{\boldsymbol #1}} 

\newcommand{\@diff}   [4]{\dfrac{#4#3#1}{#4#2#3}}
\newcommand{\diff}    [2]{\@diff{#1}{#2}{}{\dd}}
\newcommand{\pdiff}   [2]{\@diff{#1}{#2}{}\partial}
\newcommand{\fdiff}   [2]{\@diff{#1}{#2}{}{\delta}}
\newcommand{\ndiffnum}[3]{\@diff{#1}{#2}{^#3}\dd}
\newcommand{\npdiff}  [3]{\@diff{#1}{#2}{^#3}\partial}
\newcommand{\@difftwo}[4]{\dfrac{#4^2#1}{#4#2\,#4#3}}
\newcommand{\difftwo} [3]{\@difftwo{#1}{#2}{#3}{\dd}}
\newcommand{\pdifftwo}[3]{\@difftwo{#1}{#2}{#3}{\partial}}

\newcommand{\un}[1]{{\mathrm{\,#1}}} 
\newcommand{\TeV}{\un{TeV}}
\newcommand{\GeV}{\un{GeV}}
\newcommand{\MeV}{\un{MeV}}
\newcommand{\keV}{\un{keV}}
\newcommand{\invfb}{\un{fb^{-1}}}

\def\@xxxEV{\@ifnextchar-{\@xxxEV@minus}{\@xxxEV@plus}}
\def\@xxxEV@plus#1#2{%
  \ifnum{#1=0}{}\else\ifnum{#1=1}{10}\else {10^#1}\fi\fi #2}
\def\@xxxEV@minus#1#2 {10^{-#1}{\rm\,#2}}

\newcommand{\TEV}[1]{\@xxxEV{#1}{\TeV}}
\newcommand{\GEV}[1]{\@xxxEV{#1}{\GeV}}
\newcommand{\MEV}[1]{\@xxxEV{#1}{\MeV}}
\newcommand{\KEV}[1]{\@xxxEV{#1}{\keV}}

\def\EE{\@ifnextchar-{\@@EE}{\@EE}}
\def\@EE#1{\ifnum#1=1\times10\else\times10^{#1}\fi}
\def\@@EE#1#2{\!\times\!10^{-#2}}

\def\T{\@ifnextchar^{\T@u}{\@ifnextchar_{\T@d}{}}}
\def\T@u^#1{{^{#1}}\T}
\def\T@d_#1{{_{#1}}\T}

\let\lsim\lesssim

\newcommand{\s}[1]{_\mathrm{#1}}    

\newcommand{\suprm}[1]{^\mathrm{#1}} 

\newcommand{\pT}{p\s T}
\newcommand{\MET}{\slashed E\s T}

\newcommand{\@authornote}[2]{{\def\thefootnote{\fnsymbol{footnote}}\setcounter{footnote}{#1}#2\setcounter{footnote}{0}}}
\newcommand{\authornotemark}[1]{\@authornote#1{\addtocounter{footnote}{-1}\footnotemark}}
\newcommand{\authornotetext}[2]{\@authornote#1{\footnotetext{#2}}}

\makeatother 

\begin{document}

\begin{titlepage}

\begin{flushright}
UT--12--41\\
IPMU 12--0223
\end{flushright}

\vskip 3cm
\begin{center}
{\Large \bf
Gauge Mediation Models with Vectorlike Matters
\\
 at the LHC
}
\vskip 1.2cm

Motoi Endo$^{(a,b)}$,
Koichi Hamaguchi$^{(a,b)}$,
Kazuya Ishikawa$^{(a)}$,\\
Sho Iwamoto$^{(a)}$\authornotemark{1},
Norimi Yokozaki$^{(b)}$\authornotemark{1}
\vskip 0.9cm
\authornotetext{1}{Research Fellow of the Japan Society for the Promotion of Science}

{\it $^{(a)}$ Department of Physics, University of Tokyo,
   Tokyo 113--0033, Japan
\par
$^{(b)}$ Kavli Institute for the Physics and Mathematics of the Universe (Kavli IPMU), \\
University of Tokyo, Chiba 277--8583, Japan
}

\vskip 3cm

\abstract{
Gauge mediation model with vectorlike matters (V-GMSB) is one of the few viable SUSY models that explains the 126\,GeV Higgs boson mass and the muon anomalous magnetic moment simultaneously.
We explore exclusion bounds on V-GMSB model from latest LHC SUSY searches.
}
\end{center}
\end{titlepage}

\setcounter{page}{2}

\section{Introduction}
\label{sec:introduction}
A new particle which can be interpreted as the Standard Model (SM) Higgs boson with the mass of around $126\GeV$ has been discovered at both the ATLAS and CMS experiments~\cite{aad:2012gk,Chatrchyan:2012gu}.
The discovery has a great impact on the supersymmetric models and supersymmetry (SUSY) breaking scenarios, since the $126\GeV$ Higgs boson requires large radiative corrections from top-stop loops within the minimal supersymmetric standard model (MSSM). The Higgs boson mass of $126\GeV$ is explained either by heavy scalar tops of $\mathcal{O}(10-1000)\TeV$~\cite{heavy_stops} or by a large left-right mixing of the scalar tops~\cite{Okada:1990gg}~\footnote{Models with a large left-right mixing of the stops are constrained by the experimental bound from the branching ratio of the $b \to s\gamma$ decay. See, e.g., Ref.~\cite{Endo:2011gy} for the case of mSUGRA models.}.
A simple and plausible possibility is that SUSY particles are as heavy as $\mathcal{O}(10-1000)\TeV$, which can ameliorate the Polonyi/moduli problem~\cite{Polonyi} and SUSY CP/flavor problems, though the setup is disfavored by the conventional naturalness argument.

Besides the naturalness argument, light superparticles of $\mathcal{O}(100-1000)\GeV$ are suggested by the muon anomalous magnetic moment ($g-2$). The precise measurement of the muon $g-2$ at the Brookhaven E821 experiment~\cite{g-2_bnl2010} shows a deviation from the SM prediction after the latest improvements on evaluations of the hadronic vacuum polarization contributions~\cite{g-2_hadronic}, 
\begin{eqnarray}
  \delta a_{\mu} \equiv a_{\mu}^{\rm EXP} - a_{\mu}^{\rm SM}=(26.1 \pm 8.0) \times 10^{-10}.
  \label{eq:muon_g-2}
\end{eqnarray}
This corresponds to more than three sigma discrepancy, where the hadronic light-by-light contribution referred to Ref.~\cite{Prades:2009tw}.
Although there are discussions on evaluations of the hadronic light-by-light contribution (see e.g., \cite{deRafael:2012cg} for a recent review), they are expected to be settled in future, e.g., by lattice calculations~\cite{g-2_lattice}.
A large uncertainty in the hadronic vacuum polarization contributions would be reduced, particularly if a disagreement among the experimental data for the $2\pi$-channel region could be resolved~\cite{g-2_hadronic,Benayoun:2012wc}.
Moreover, new experiments are prepared to improve the precision of the experimental value at Fermilab~\cite{LeeRoberts:2011zz} and J-PARC~\cite{Iinuma:2011zz}.
Consequently, although the discrepancy \eqref{eq:muon_g-2} is not a definitive evidence of physics beyond the SM, it is interesting to explore new physics models in light of the muon $g-2$ as well as the Higgs boson mass of $126\GeV$. In the MSSM, the above deviation of the muon $g-2$ suggests sleptons, charginos and neutralinos at $\mathcal{O}(100)\GeV$ and large $\tan\beta$~\cite{SUSY_gminus2}.

For light superparticles of $\mathcal{O}(100-1000)$ GeV, gauge mediated SUSY breaking (GMSB) models  are attractive from phenomenological points of view~\cite{Giudice:1998bp}. They are free from the SUSY flavor problem, and dangerous CP violating phases do not appear in scalar and gaugino masses. 
However, the models have difficulty in explaining the Higgs boson mass of $126\GeV$ as long as superparticle masses are $\mathcal{O}(100-1000)\GeV$, because the left-right mixing of the stops is too small to enhance the Higgs boson mass.
As a result, GMSB scenario requires an extension to explain both the Higgs boson mass of $126\GeV$ and the muon $g-2$.

Recently, various extensions of the gauge mediation models have been studied for this purpose: 
the vectorlike-matter-assisted GMSB (V-GMSB)~\cite{Endo:2011mc,Endo:2011xq}~\footnote{The V-GMSB model was also studied in Refs.~\cite{Evans:2011uq,Martin:2012dg}, where the muon $g-2$ was not taken into account.},
additional ${\rm U}(1)'$ gauge symmetry imposed on the Higgs fields~\cite{Endo:2011gy}, 
a large left-right mixing of the stops induced by introducing a Higgs-messenger mixing~\cite{EIY}, 
and splitting $F$-terms between colored and non-colored messengers~\cite{Ibe:2012qu} (see also \cite{Sato:2012bf}).
Among the models, we will study the V-GMSB model in this paper. 
The model can enhance the Higgs boson mass easily by extra top-like matter multiplets in a similar manner to the top-stop contributions~\cite{OkadaMoroi1992vectorlike,Babu:2008ge,Martin:2009bg}.
It has been shown that the model can explain both the Higgs boson mass of $126\GeV$ and the muon $g-2$ simultaneously in a vast range of parameter space~\cite{Endo:2011mc,Endo:2011xq}~\footnote{Also, the model realizes the perturbative gauge coupling unification.}.

In this paper, we explore LHC status of V-GMSB in light of the latest SUSY search results of the ATLAS and CMS experiments. SUSY signatures of the V-GMSB model at the LHC have been briefly discussed in Refs.~\cite{Endo:2011xq,Martin:2012dg}.
In this paper, the latest LHC results are applied to constrain the parameter space of V-GMSB, and we will show whether the region where the muon $g-2$ is explained is still valid or not. Since the SUSY contributions to the muon $g-2$ are suppressed as the superparticles are heavier, the V-GMSB model has an upper bound on the soft mass scale, e.g., on the gluino and squark masses. Thus, the model can be tested by direct searches for the superparticles. We will also improve evaluation of the superparticle mass spectrum. Relevant corrections from the vectorlike matters will be taken into account. It will be shown that the parameter regions of the gluino mass of $<1000-1100\GeV$ are excluded in the case when the neutralino is the next-lightest-superparticle (NLSP), and more severe constraint is obtained for the stau NLSP. Consequently, it will be found that the whole regions where the muon $g-2$ is explained at the $1\sigma$ level 
are excluded by the LHC at 95\% confidence level (CL). 
The region where the muon $g-2$ is explained at $2\sigma$ is still viable.

This paper is organized as follows. In section \ref{sec:model}, we will introduce the V-GMSB model, and discuss the Higgs boson mass, the mass spectrum, the muon $g-2$, and constraints on the model.
In section \ref{sec:lhc_limit}, analysis methods to study the collider searches for the superparticles will be described, and the latest LHC results will be applied.
The main results are summarized in Fig.~\ref{fig:results1} and \ref{fig:results2}.
Section \ref{sec:discussion} is devoted to conclusion and discussion.

\section{V-GMSB Model}
\label{sec:model}
\subsection{V-GMSB model and Higgs boson mass}

Let us explain the vectorlike-matter-assisted GMSB, or V-GMSB, briefly. The model with the GMSB boundary condition introduces a pair of the vectorlike matters in the representations of ${\bf 10}=(Q', \bar{U}', \bar{E}')$ and $\overline{{\bf 10}}=({\bar Q}', U', E')$ of the SU(5) GUT gauge group. These extra matters generally have Yukawa couplings with the Higgs boson supermultiplets. The superpotential and soft SUSY breaking terms (below the SUSY-breaking scale) are respectively given by
\begin{equation}
W = W_{\rm MSSM} + Y' Q' H_u \bar{U}' + Y'' \bar{Q}' H_d U' + M_{Q'} Q' \bar{Q}' + M_{U'} U' \bar{U}' + M_{E'} E' \bar{E}',
\end{equation}
and
\begin{align}
\begin{split}
 -\mathcal{L}_{\rm soft}
 &=
 -\mathcal{L}_{\rm soft}^{\rm MSSM}
\\&\qquad
 + m_{{Q}'}^2 |Q'|^2 + m_{{\bar{Q}}'}^2 |\bar{Q}'|^2 + m_{{U}'}^2 |U'|^2 + m_{{\bar{U}}'}^2 |\bar{U}'|^2
 + m^2_{E'}|E'|^2 + m^2_{\bar E'}|\bar E'|^2
\\&\qquad
 + \left(A' Y'  Q' H_u \bar{U}' + A'' Y'' \bar{Q}' H_d U' + \text{h.c.}\right)
\\&\qquad
 + \left(B_{Q'} M_{Q'} Q' \bar{Q}' + B_{U'} M_{U'} U' \bar{U}' + B_{E'} M_{E'} E' \bar{E}' + \text{h.c.}\right).
\end{split}\end{align}
The up-type (down-type) Higgs field is denoted as $H_u$ ($H_d$). 
We assume that the mixings between the vectorlike fields and MSSM fields are suppressed, though there must be small but finite mixings in order to avoid cosmological difficulties due to stable exotic particles. Such mixings induce decays of the extra matters, which will be briefly discussed in Sec.~\ref{sec:lhc_limit}.

The vectorlike top-like quarks, $t'_1$ and $t'_2$, and their superpartners give radiative correction via the Yukawa interaction $Y'Q'H_u\bar U'$ to the Higgs boson mass in a similar manner to the top-stop loop corrections.
This contribution is approximately expressed as~\cite{Martin:2009bg}
\begin{equation}
 \begin{split}
\label{eq:Higgs-mass}
\Delta m_h^2 &\approx
\frac{3Y'^4 v^2}{4\pi^2} \left[\ln \frac{M_S^2}{M_F^2} - \frac{1}{6}\left(1-\frac{M_F^2}{M_S^2}\right)\left(5-\frac{M_F^2}{M_S^2}\right) + \frac{A'^2}{M_S^2} \left(1-\frac{M_F^2}{3M_S^2} \right)-\frac{1}{12} \frac{A'^4}{M_S^4}\right],
\end{split}
\end{equation}
where the approximations of
$M_{Q'}=M_{U'}(\equiv M_F)$, $m^2_{Q'}=m^2_{\bar Q'}=m^2_{U'}=m^2_{\bar U'}(\equiv M_S^2- M_F^2)$ and $B_{Q'}=B_{U'}=0$ have been applied.
In the numerical analysis, we include the full one-loop contributions.

The radiative correction \eqref{eq:Higgs-mass} becomes effective when the Yukawa coupling $Y'$ is close to unity. 
Unless $Y'$ is suppressed at a high scale, e.g., at the GUT scale, this is likely to be realized at low-energy scale, since it has a quasi-infrared fixed point at $Y' \simeq 1$~\cite{Martin:2009bg}.
On the contrary, the other Yukawa coupling $Y''$ should be suppressed~\footnote{%
The suppression of $Y''$ can be explained in a model with the PQ symmetry~\cite{Nakayama:2012zc},
where $M_{Q', U', E'}\sim \mu\sim \mathcal{O}(1)$ TeV is also related to the PQ-symmetry breaking scale.
},
since the interaction $Y''\bar{Q}' H_d U'$ decreases the Higgs boson mass~\cite{Endo:2011xq}. 
If the couplings take the values of $(Y',Y'')\sim(1,0)$, the Higgs boson mass of 126\,GeV is explained with
$M_{Q'}, M_{U'}\sim{\mathcal O}(100-1000)\GeV$ and $M_S\sim {\mathcal O}(1000)\GeV$.

In the V-GMSB model, the soft SUSY-breaking terms are induced radiatively via messenger particles as in the ordinary GMSB scenario. 
The superpotential of the messenger sector is
\begin{eqnarray}
W_{\rm mess} = ({\mathcal M}_D + {\mathcal F}_D \theta^2) \Psi_D \bar{\Psi}_D + ({\mathcal M}_L + {\mathcal F}_L \theta^2) \Psi_L \bar{\Psi}_L,
\end{eqnarray}
where ${\mathcal M}_D \, ({\mathcal M}_L)$ is the supersymmetric mass, and ${\mathcal F}_D\, ({\mathcal F}_L)$ is the SUSY breaking $F$-term of the colored (non-colored) messengers. In the following, 
only a pair of the ${\vc 5}+\overline{\vc 5}$ representation is included, which corresponds to the messenger number $N_{\rm mess}=1$.
The messenger scale is taken as $M_{\rm mess} \equiv {\mathcal M}_D = {\mathcal M}_L$, and the SUSY breaking F-terms are as ${\mathcal F}_D={\mathcal F}_L$. Thus, the soft SUSY-breaking parameters are characterized by the two parameters, ${\mathcal M}_{\rm mess}$ and $\Lambda\equiv{\mathcal F}_D/{\mathcal M}_D={\mathcal F}_L/{\mathcal M}_L$, at the messenger scale~\footnote{
The relation, ${\mathcal M}_D={\mathcal M}_L$ and ${\mathcal F}_D={\mathcal F}_L$, does not hold generally at the messenger scale due to the RG evolution, even if we take the universal condition, ${\mathcal M}_D={\mathcal M}_L$ and ${\mathcal F}_D={\mathcal F}_L$, at the GUT scale. However, $\Lambda$ is almost unchanged during the RG running, and hence, the soft mass parameters are not significantly changed.
}.

One of the characteristic features of the V-GMSB model is a size of the Higgsino mass parameter $\mu$ at the SUSY scale. 
It becomes very large compared to the usual GMSB models. 
This is because the vectorlike matters contribute to the renormalization group equation (RGE) of the up-type Higgs mass squared, which is approximately given by
\begin{eqnarray}
\frac{d m_{H_u}^2}{d \ln Q} 
\simeq 
\left.\frac{d m_{H_u}^2}{d \ln Q}\right|_{\rm MSSM}
+
\frac{3}{8 \pi^2} Y'^2 ( m_{Q'}^2 + m_{\bar{U}'}^2 + m_{H_u}^2 + |A'|^2),
\label{eq:RG_mhu}
\end{eqnarray}
where $Q$ is the renormalization scale. The complete formula at the two-loop level is found in Ref.~\cite{Endo:2011mc}.
The second term comes from the vectorlike matters, which gives a large negative contribution to the up-type Higgs mass squared $m_{H_u}^2$.
It is as large as the stop contribution for $Y' \simeq 1$.
Hence, the electroweak symmetry breaking conditions require a large value of $\mu$.
Consequently, the next-to-lightest neutralino $\tilde\chi_2^0$ and the lightest chargino $\tilde\chi_1^\pm$ tend to consist of the Wino, and the mass of the lighter stau is likely to be small, because the left-right mixing of the stau mass matrix, which is proportional to $\mu \tan\beta$, becomes large. Especially when $\tan\beta$ is large,  the lighter stau becomes the NLSP.  These features can be found in the mass spectrum discussed below.

\subsection{Mass spectrum}

 \begin{figure}[t]
 \begin{center}
  \includegraphics[height=205pt]{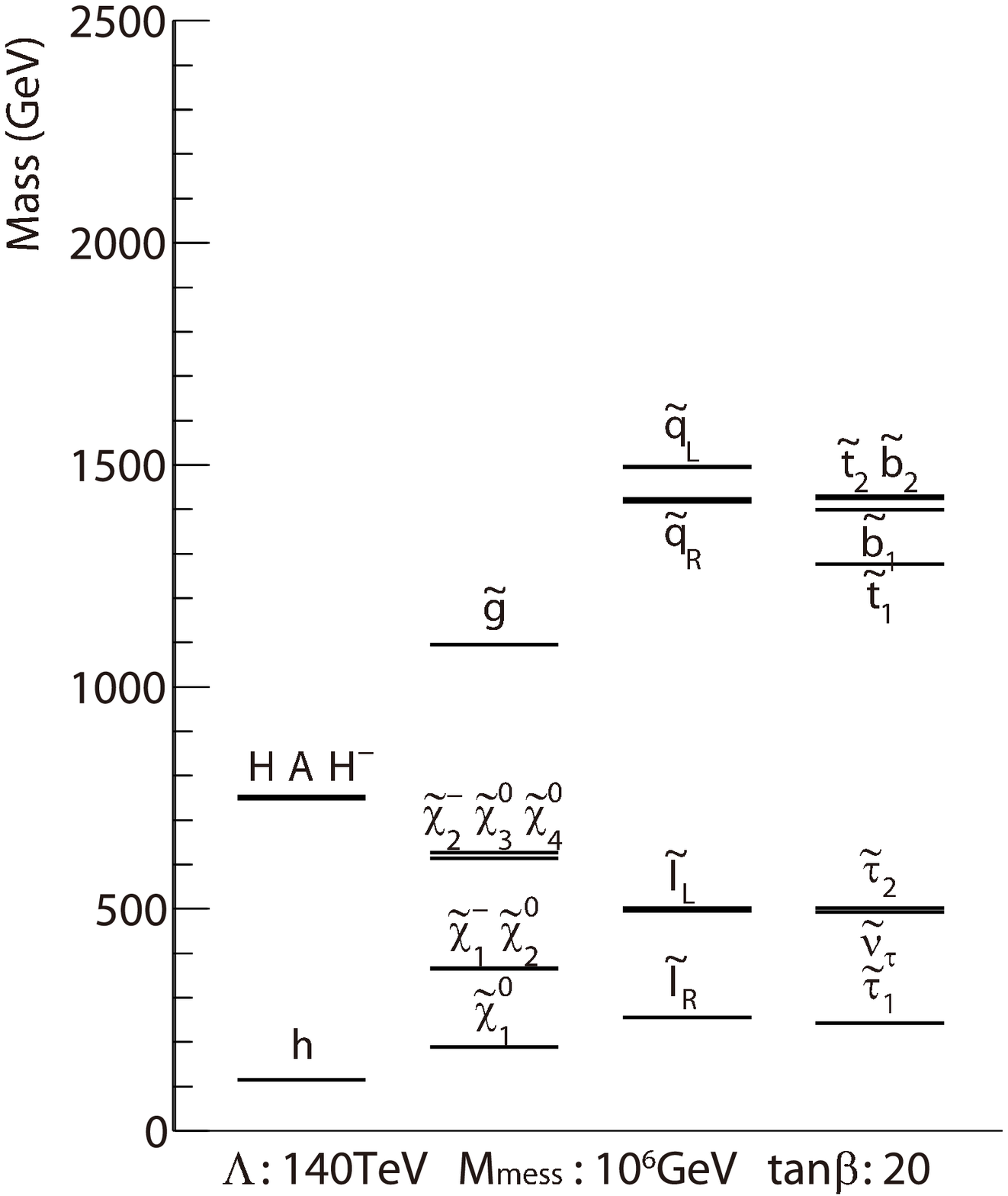}
  \hspace{10pt}
  \includegraphics[height=205pt]{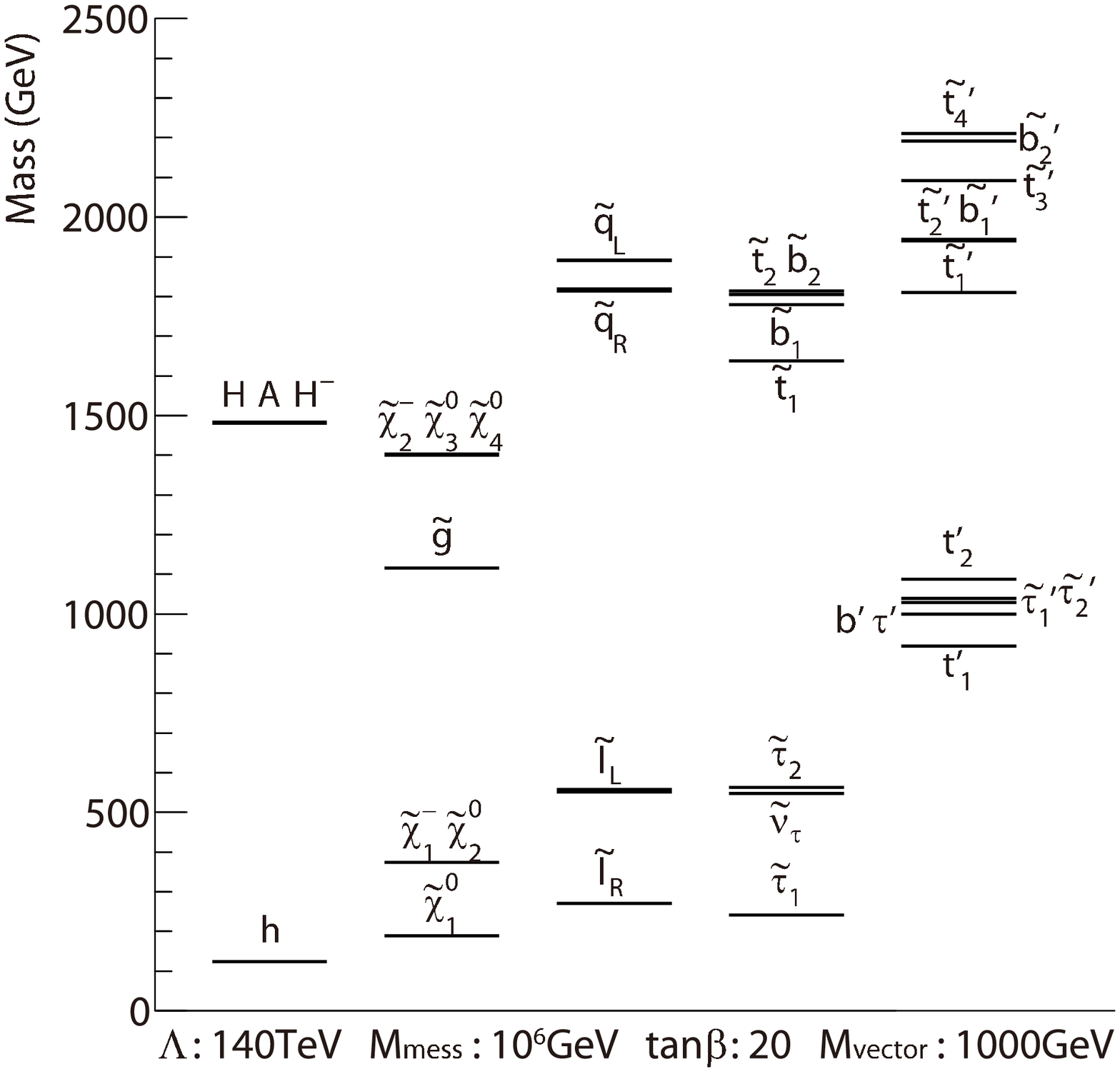}
  \hspace{10pt}
  \caption{%
The mass spectra of the GMSB model (left) and the V-GMSB model (right). The GMSB parameters are $(\Lambda, M\s{mess},\tan\beta,N_{\rm mess})=(140\TeV,10^6\GeV,20,1)$ in both cases.
The SUSY-invariant mass of vectorlike fields are set as $M_{Q'}=M_{U'}=M_{E'}=1\TeV$ for the V-GMSB model.
The masses of vectorlike fermions (scalars) are labelled by
$\tau'$, $b'$, and $t'_{1,2}$ 
($\widetilde{\tau}_{1,2}'$, $\widetilde{b}_{1,2}'$, and $\widetilde{t}_{1,2,3,4}'$), respectively.
}
  \label{fig:spectrum}
 \end{center}
 \end{figure}

In this subsection, we discuss the mass spectrum of the V-GMSB.
In the numerical analysis we utilized {\tt SOFTSUSY 3.3}~\cite{softsusy} and {\tt FeynHiggs 2.9}~\cite{FeynHiggs} to calculate the mass spectrum of the SUSY particles and the Higgs sector. They are modified to take the effects from the extra vectorlike matters into account. 

A typical mass spectrum is displayed in Fig.~\ref{fig:spectrum}, where the V-GMSB result is compared to that of the ordinary GMSB model.
Here and hereafter, we assume a common SUSY-invariant mass for the vectorlike fields, $M_V=M_{Q'}=M_{U'}$ ($=M_{E'}$). As mentioned in the previous subsection, the heavier chargino $\tilde{\chi}^\pm_2$ and the heavier two neutralinos $\tilde{\chi}^0_{3,4}$ mainly consist of the Higgsino, and are much heavier than the ordinary case. Also, it is found that the squarks become heavier.
This is because the beta function of the strong gauge coupling constant is zero at the one-loop level, 
and the coupling
stays large during the RGE.
On the contrary, a ratio of the gaugino masses are less affected by the extra matters, since the ratios of $M_i/\alpha_i$, where $M_i$ are the gaugino masses, are fixed during the RGE at the one-loop level.
The masses of the scalar vectorlike matters are close to the top and bottom squarks up to the SUSY-invariant mass, because they have the same quantum numbers as corresponding squarks. The masses of the fermionic vectorlike matters are determined by the SUSY-invariant mass with the electroweak symmetry breaking corrections for $t'$.

Let us explain how the mass spectrum is evaluated in {\tt SOFTSUSY} and what kinds of modifications are applied.
First of all, the program estimates the SM gauge couplings $\alpha_a(M_Z)$ and Yukawa couplings $Y_i(M_Z)$ at the scale of the $Z$-boson mass $M_Z$~\cite{Pierce:1996zz}.
They are evolved upwards from $M_Z$ to the messenger scale $M_{\rm mess}$ by solving the V-GMSB RGE, namely, including the contributions from the vectorlike matter supermultiplets.
In the numerical calculations, the {\tt SOFTSUSY} package is modified to include the two-loop RGE of the V-GMSB model, which can be found in~\cite{Endo:2011mc}.
It is also important to include threshold/self-energy corrections of vectorlike matters to the gauge coupling constants at the $M_Z$ scale, since they can give $\sim 10$\% contributions especially to the strong coupling constant.
They also affect the gaugino masses and the scalar masses squared generated at $M_{\rm mess}$, which are respectively proportional to $g^2$ and $g^4$.

Next, the soft SUSY breaking parameters are provided by the messenger loops at $M_{\rm mess}$.
The extra Yukawa couplings are also set at the scale, which are $Y'(M_{\rm mess}) = 1$ and $Y''(M_{\rm mess}) = 0$.
The gauge, Yukawa and soft parameters are evolved by the V-GMSB RGE down to the SUSY scale, $M_{\rm SUSY}$, which is determined by the stop masses as $M_{\rm SUSY}=\sqrt{m_{\tilde{t}_1}m_{\tilde{t}_2}}$.

The masses of the superparticles are evaluated including the whole one-loop corrections to the self-energies within the MSSM.
The pole mass of the gluino also receives a correction from vectorlike matter loops.
However, this turns out to be around a few GeV, since the masses of 
the vectorlike matters are close to $M_{\rm SUSY}$, and their contribution to the self-energy are relatively small.
Similarly, contributions of the vectorlike fields to the electroweak gaugino masses are safely neglected.

Then, the Higgs potential is investigated, which determines the Higgs boson mass and the $\mu$ parameter. 
The MSSM part of the Higgs boson mass is evaluated at the two-loop level by the {\tt FeynHiggs} package~\cite{FeynHiggs}, while that from the vectorlike matters is estimated with the one-loop level effective potential~\cite{Martin:2009bg}, where all one-loop contributions from the vectorlike matters are taken into account.
The two contributions to the Higgs boson mass are added in quadrature.
The two-loop contribution of the vectorlike matters can shift the Higgs boson mass by $\sim 1-10\GeV$, study for which is reserved for future work. On the other hand, the $\mu$ parameter is less affected by the vectorlike matters except for the RG evolution~\eqref{eq:RG_mhu}, because the tree-level contribution to $\mu$ is quite large in the electroweak symmetry breaking condition.

There are several updates compared to the previous works in Ref.~\cite{Endo:2011mc,Endo:2011xq}, where the {\tt SuSpect2}~\cite{SuSpect2} package was used to estimate the superparticle masses.
In the old analyses the threshold/self-energy corrections to the gauge coupling constants from vectorlike matters were not included, which yield $\sim 10$\% difference in the gauge couplings and the soft parameters for the GMSB scenario.
Also, in Ref.~\cite{Endo:2011mc,Endo:2011xq}, the $\mu$ term contribution was discarded for simplicity in the estimation of the Higgs boson mass. In the present work, those contributions are included, which shift the Higgs boson mass by a few GeV.
In addition, the self-energy correction from the MSSM particles to the stau mass is not fully included in {\tt SuSpect2}.
In the V-GMSB model, since the left-right mixing is relatively large, the lighter stau mass receives a sizable correction from the self-energy correction, leading to an $\mathcal{O}(10)\,{\rm GeV}$ shift for a large $\tan\beta$.
The mass spectra  in the current analysis differs from those in Refs.~\cite{Endo:2011mc,Endo:2011xq} as well as Ref.~\cite{Martin:2012dg}.

\subsection{Muon $g-2$}

In the V-GMSB model, the leading contributions to the muon $g-2$ come from the smuon-neutralino and/or muon sneutrino-chargino loops as~\cite{SUSY_gminus2}
\begin{eqnarray}
  \Delta a_\mu({\rm chargino}) 
  &\simeq& \frac{\alpha_2m_\mu^2}{8\pi m_{\rm soft}^2} {\rm sgn}(\mu M_2) \tan\beta, \nonumber\\
  \Delta a_\mu({\rm neutralino}) 
  &\simeq& \frac{\alpha_Ym_\mu^2}{24\pi m_{\rm soft}^2} {\rm sgn}(\mu M_1) \tan\beta + \ldots,
  \label{eq:g-2-SUSY}
\end{eqnarray}
where $m_{\rm soft}$ is a typical scale of the soft parameters and the $\mu$ parameter.
It is evident that the SUSY contributions are enhanced when $\tan\beta$ is large and $m_{\rm soft}$ is small. 
In the following, $\mu > 0$ is chosen, since the discrepancy \eqref{eq:muon_g-2} is positive.
The vectorlike matters contribution to muon $g-2$ is negligibly small.
In the numerical analysis,
 the SUSY contributions to the muon $g-2$ is evaluated by {\tt FeynHiggs}, which includes two-loop contributions of the MSSM particles.

\subsection{Vacuum stability}
\label{sec:vac}

One of the most severe constraints on the V-GMSB model is the vacuum stability bound~\cite{Endo:2012rd}. When $\tan\beta$ is large, the large trilinear coupling of the staus and the Higgs boson can generate charge breaking global minima and destabilize the electroweak symmetry breaking vacuum~\cite{Rattazzi:1996fb}. Given $M_{\rm mess}$ and $\Lambda$, $\tan\beta$ is tightly bounded from above by the condition that the lifetime of the electroweak symmetry breaking vacuum is longer than the age of the universe.
In Ref.~\cite{Endo:2012rd}, a fitting formula given in Ref.~\cite{Hisano:2010re} was utilized. 
Recently, however, the fitting formula has been revised as follows~\cite{Hisano:2010re}:
\begin{equation}
\begin{split}
 \mu\tan\beta &<
 213.5\sqrt{m_{\tilde{L}_3} m_{\tilde{\tau}_R}} 
 -17.0 (m_{\tilde{L}_3} + m_{\tilde{\tau}_R}) 
 \\
 &\qquad
+ 4.52\times 10^{-2} {\rm GeV}^{-1} (m_{\tilde{L}_3} - m_{\tilde{\tau}_R})^2
 -1.30 \times 10^4\ {\rm GeV}, \label{eq:fitting}
\end{split}
\end{equation}
where $m_{\tilde{L}_3}$ and $m_{\tilde{\tau}_R}$ are the soft masses for the left- and right-handed stau, respectively. 
Moreover, also very recently, the vacuum stability bound has been reanalyzed including effects of a radiatively-corrected tau Yukawa coupling~\cite{Carena:2012mw}.
The results in Ref.~\cite{Carena:2012mw}, however, cannot be directly applied to the V-GMSB model, because the masses and the mixing angle of the staus in our model are different from those in Ref.~\cite{Carena:2012mw}.
In addition, the thermal transition may change the bound~\cite{staureheating}.
In this paper we show the bound of Eq.~\eqref{eq:fitting}, and also draw a bound which is weakened by 10\% as a reference.


\subsection{Numerical Result}

\begin{figure}[t!]
\begin{center}
\begin{tabular}{cc}
\includegraphics[width=0.41\linewidth]{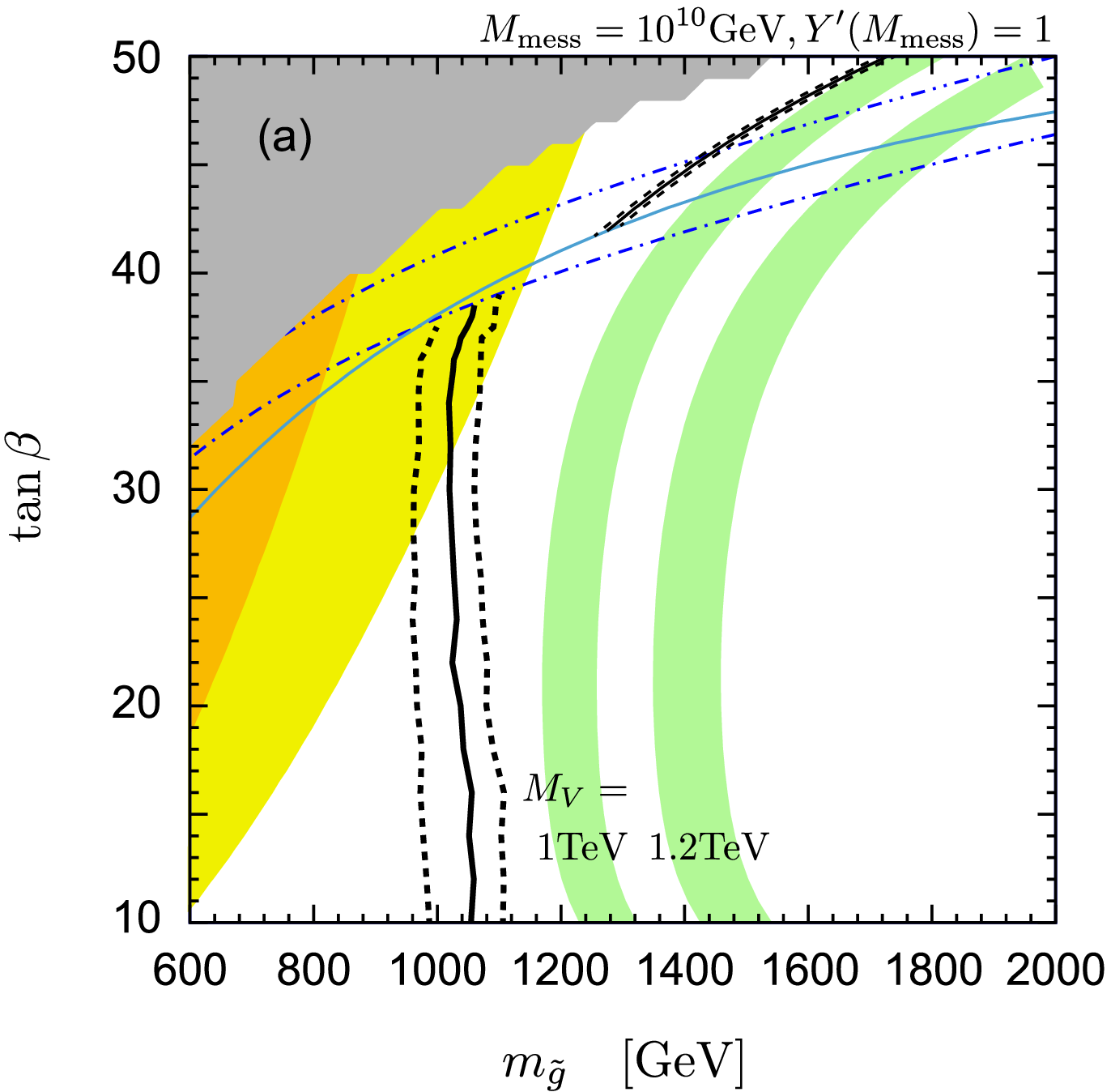}  &
\includegraphics[width=0.41\textwidth]{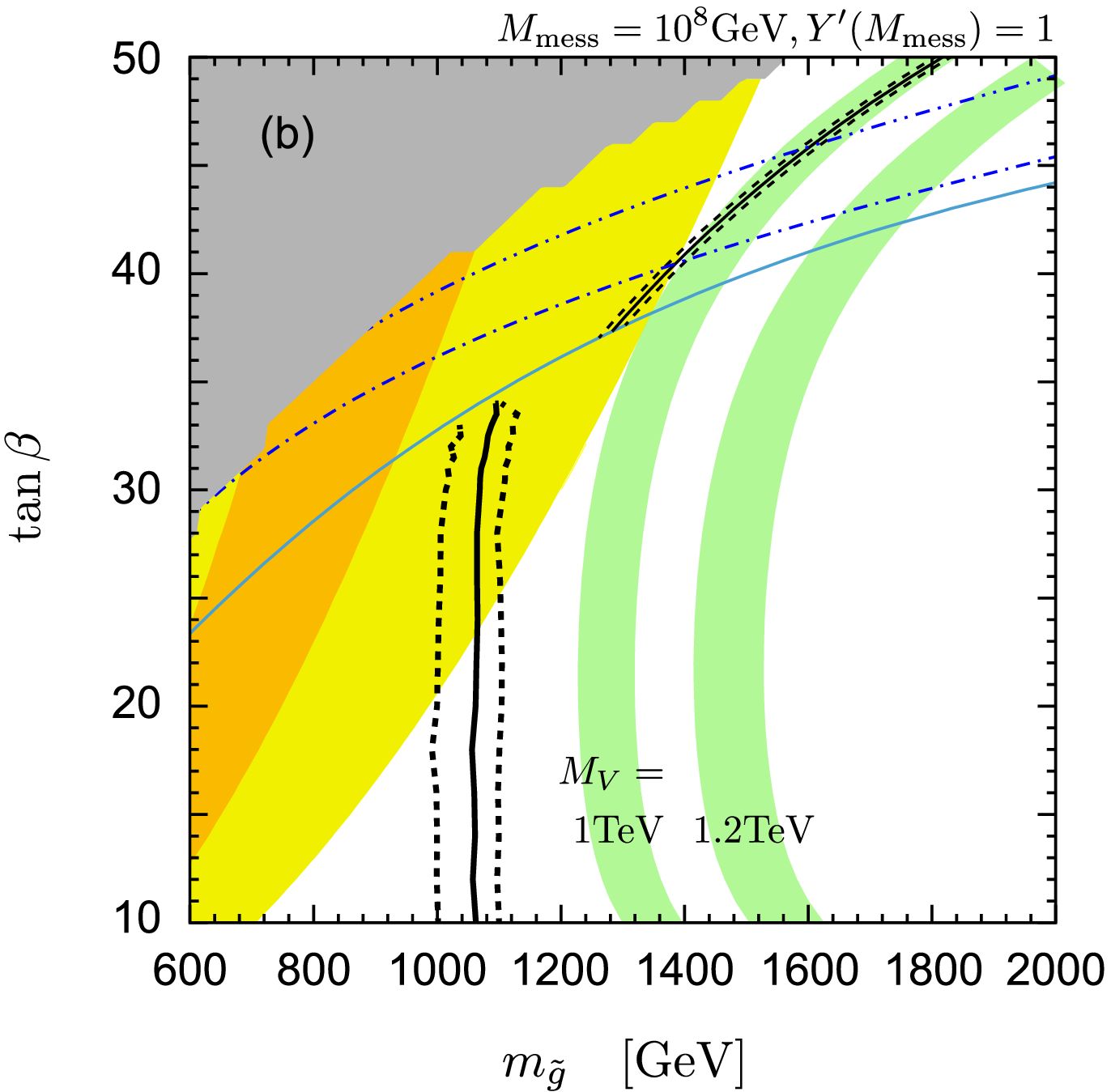}  \\
\includegraphics[width=0.41\textwidth]{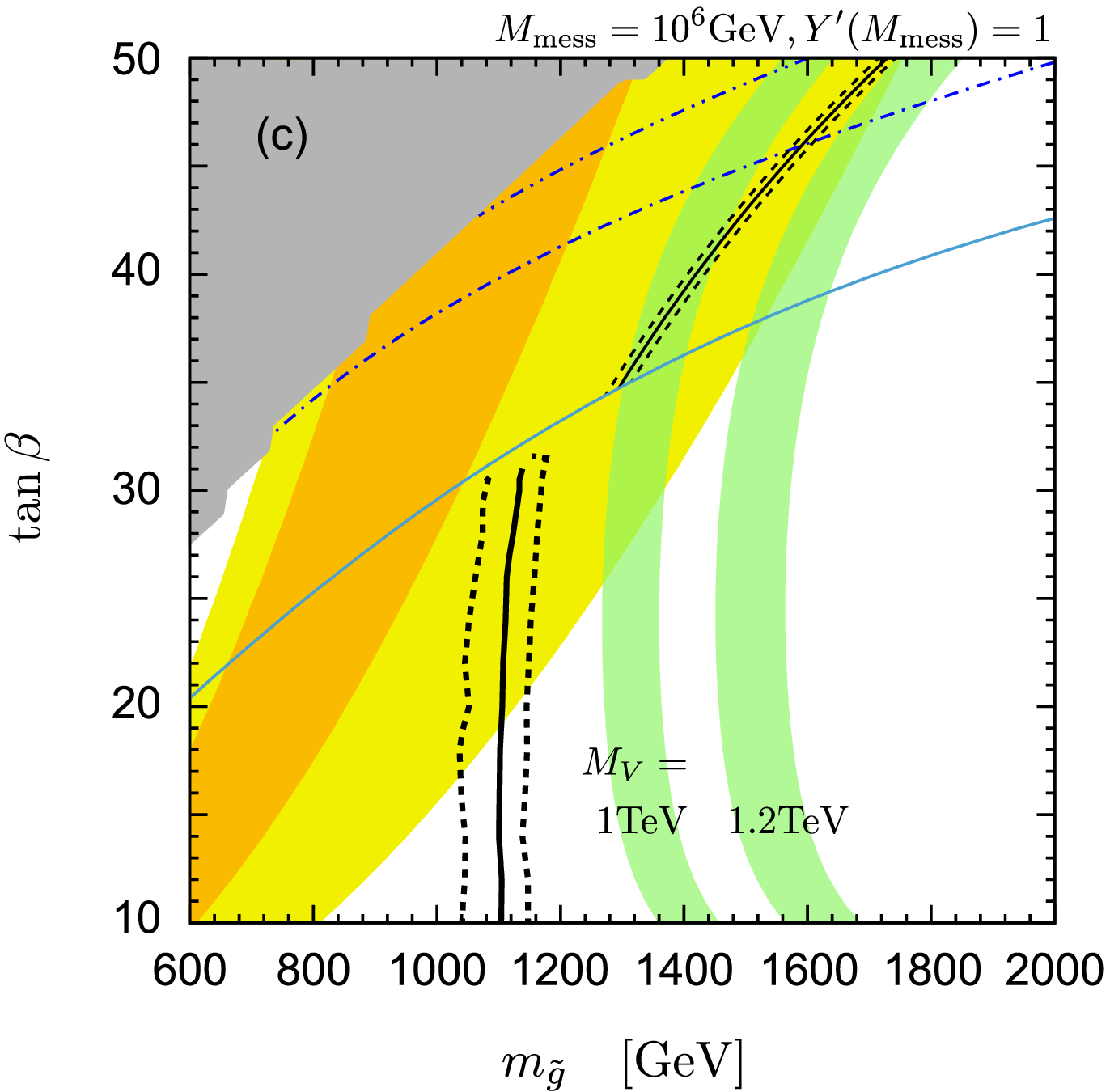} &
\includegraphics[width=0.41\textwidth]{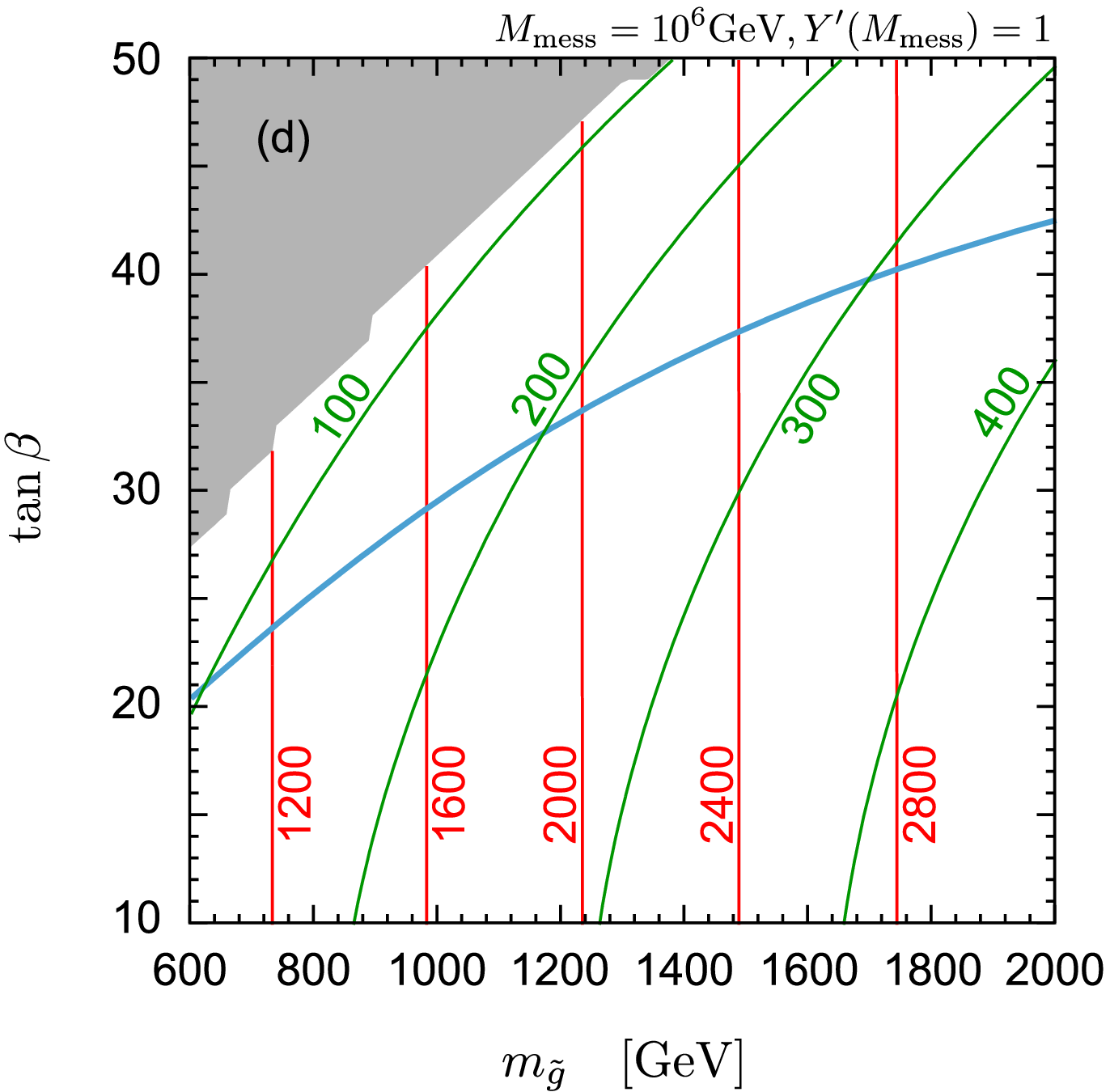}
\end{tabular}
\end{center}
\caption{
Contours of the Higgs boson mass, the muon $g-2$ and the LHC constraints in the V-GMSB model
are shown for 
(a) $M_{\rm mess}=10^{10}\GeV$,
(b) $10^{8}\GeV$, and 
(c) $10^{6}\GeV$. In the green bands, the SM-like Higgs boson mass is $125-126\GeV$ for the masses of the vectorlike matter, 
$M_V=1\TeV$ and $1.2\TeV$. In the yellow (orange) region, the muon $g-2$ discrepancy is within $2\sigma$ ($1\sigma$). Below the light blue lines, the NLSP is the neutralino, whereas the NLSP is the stau above them. 
The LHC constraints are shown by the black solid lines without theoretical uncertainties, where the regions left to the lines are excluded. 
The black dashed lines include theoretical uncertainties: 
35\% error of the production cross section is adopted for the neutralino NLSP, and 2\% error in terms of the stau mass is for the stau NLSP. 
The regions above the blue dot-dashed lines are constrained by the vacuum stability condition, \eqref{eq:fitting}, and the bound which is weakened by 10\% of \eqref{eq:fitting} is drawn by the blue double-dotted long-dashed lines. 
The gray shaded regions are excluded by the failure of the {\tt SOFTSUSY} calculation. 
In figure (d), masses of the stau and the lightest squark among the first two generations are shown respectively by green and red lines for $M_{\rm mess}=10^6\GeV$.
}
\label{fig:results1}
\end{figure}

Before moving on to the LHC constraints on the V-GMSB model, let us show the numerical results. The current status of the V-GMSB model is summarized in Fig.~\ref{fig:results1}. (See also Fig.~\ref{fig:results2}).

In Fig.~\ref{fig:results1}, contours of the Higgs boson mass and the muon $g-2$ are drawn in the $(m_{\tilde{g}}, \tan\beta)$ plane
for the messenger scales, (a) $M_{\rm mess}=10^{10}\GeV$, (b) $10^{8}\GeV$, and (c) $10^{6}\GeV$. Here, $m_{\tilde{g}}$ is the gluino pole mass, and $\tan\beta$ is a value at the scale, $Q=M_Z$. In this paper, we use the strong coupling constant of $\alpha_s(M_Z)=0.1184$, and the top pole mass of $m_t=173.5$GeV. 
As discussed in Sec.~\ref{sec:vac}, the vacuum stability bound is imposed by Eq.~\eqref{eq:fitting}. In the figure, it is drawn by the blue dot-dashed line. The blue double-dotted long-dashed line corresponds to the vacuum stability condition which is weakened by 10\%.
The black solid and dashed lines show the LHC constraints, which we will discuss in detail in Sec.~\ref{sec:lhc_limit}.
In the gray shaded region, {\tt SOFTSUSY} fails to calculate the stau mass. Note that such a parameter region is experimentally excluded, as long as the NLSP is long-lived.


In the green bands in Figs.~\ref{fig:results1} (a)-(c), the SM-like Higgs boson mass takes a value of $125-126\GeV$ for reference values of SUSY-invariant mass of the vectorlike matter, $M_V=1\TeV$ and $1.2\TeV$. The Higgs boson mass increases as the vectorlike matter becomes lighter. The Higgs boson mass of $125-126\GeV$ can be realized in the whole parameter region of Fig.~\ref{fig:results1} by changing the vectorlike matter mass. 
For fixed $M_V$, the Higgs boson becomes lighter when the gluino mass is smaller, because the radiative corrections to the Higgs potential from (vectorlike-) top squarks decrease.
On the other hand, 
if the messenger scale is higher, the (vectorlike-) top squark masses become larger due to the RG correction, and the Higgs boson mass of $125-126\GeV$ is realized for smaller gluino mass. 

Although the Higgs boson mass depends on $M_V$, masses of the MSSM superparticles and the SUSY contributions to the muon $g-2$ are less sensitive to $M_V$. We have checked that the MSSM superparticle masses change only by $\lesssim 2$\% when $M_V$ is varied from $500\GeV$ to $1\TeV$, and the muon $g-2$ changes by less than 1\%, correspondingly. In the figures, we fix $M_V=1\TeV$ in calculating all the quantities except for the Higgs boson mass.


In the yellow (orange) region in Figs.~\ref{fig:results1} (a)-(c), the muon $g-2$ discrepancy, Eq.~\eqref{eq:muon_g-2}, is explained at the $2\sigma$ ($1\sigma$) level. As shown by Eq.~\eqref{eq:g-2-SUSY}, the region is sensitive to both the gluino mass, i.e., the soft mass scale, and $\tan\beta$. It is stressed that the soft mass scale has an upper bound in order to explain the muon $g-2$ discrepancy. Also, too large $\tan\beta$ spoils the vacuum stability condition. 
The bound becomes tighter as $M_{\rm mess}$ increases. This is because the soft scalar masses are enhanced by RGE compared to the gaugino masses, which leads to suppression of the SUSY contributions to the muon $g-2$ for a fixed value of the gluino mass.

As we shall see in the next section, the LHC search for the SUSY particles depends on the species of the NLSP. In Figs.~\ref{fig:results1} (a)-(c), the NLSP is the neutralino (the stau) below (above) the light blue line, respectively. As $M_{\rm mess}$ increases, the RG correction lifts up the scalar masses relative to the gaugino masses, and thus, the stau NLSP region becomes smaller.
For the LHC search, the masses of the gluino, squarks, and the stau are particularly important. For illustration, we show  in Fig.~\ref{fig:results1} (d) the masses of the lighter stau and the lightest squark among the first two generations by green and red lines, respectively, for $M_{\rm mess}=10^6\GeV$.
The squark mass is almost independent of $\tan\beta$, since they are governed by the strong interaction.
On the other hand, the stau mass, which determines the LHC constraint in the parameter region with the stau NLSP, depends both on the gluino mass, i.e., the soft mass scale, and $\tan\beta$ through the tau Yukawa coupling.


\section{LHC Limit}
\label{sec:lhc_limit}
Let us move on to the main topic of this paper. Constraints on the V-GMSB model from the LHC experiments will be explored.
The current status of the superparticle searches at the LHC is applied to the V-GMSB model. 

In the model, the gravitino is the LSP, and collider signatures are determined by species and the lifetime of the NLSP. 
As discussed in the previous section, the NLSP is either the neutralino or the stau in the V-GMSB model, which are realized below and above the light blue lines in Fig.~\ref{fig:results1}.
In this paper, we concentrate on the case with a quasi-stable NLSP~\footnote{
The NLSP could decay promptly if the gravitino is lighter than $\mathcal{O}(1)\keV$.
Then, collider signatures will change, such as events of di-photon with large missing energy for the neutralino NLSP case, and those of multi-taus with large missing energy for the case of the stau NLSP.
Also, small violations of the R-parity can lead to a prompt decay of the NLSP.
These cases require independent studies on LHC constraints, and are reserved for future works. See also discussion in Sec.~\ref{sec:discussion}.
}.
If a quasi-stable neutralino is the NLSP, searches for the events with no lepton, multi jets and missing energy give the most stringent constraint. On the other hand, long-lived slepton searches are the relevant search for a quasi-stable stau NLSP. These searches will be discussed in the following subsections.

In the analysis, the mass spectrum is generated with customized {\tt SOFTSUSY 3.3}~\cite{softsusy} as explained in the previous section, and passed to {\tt SUSY-HIT 1.3}~\cite{SUSYHIT} to calculate the decay pattern of the SUSY particles and the Higgs bosons.
{\tt Prospino 2.1}~\cite{Prospinoweb,ProspinoSG,ProspinoNL} is utilized to calculate the next-to-leading order cross section.
The {\tt CTEQ6L1} and {\tt CTEQ6.6M} parton distribution functions (PDFs)~\cite{PDFCTEQ6}
are used in the cross section calculation; the former is also used in Monte Carlo simulation
for the event generation.

Before proceeding to details of the LHC study, let us mention collider searches for the vectorlike quarks.
In the V-GMSB model, the Higgs boson mass puts an upper bound on the masses of the vectorlike matters~\cite{Endo:2011xq}~\footnote{%
A precise  evaluation of the upper bound on the vectorlike matter masses requires two-loop contributions of the vectorlike matters to the Higgs potential, which is beyond the scope of this paper. Note that the main results of this paper, the constraints from SUSY searches at the LHC, are almost independent of the masses of the vectorlike fields.}.
Those particles are searched for in the LHC experiments.
For example, the lighter top-like quark $t_1'$ has a lower mass bound of $m_{t'_1} > 656\GeV$ if it decays exclusively into $b W$~\cite{ATLAS:2012qe}.
However, their decay pattern crucially depends on the mixing pattern between the vectorlike matters and the MSSM matters.
$t_1'$ may have other decay channels such as $t'_1\to qZ$ and $t'_1\to qh$, and then the constraints becomes much looser~\cite{Endo:2011xq, Harigaya:2012ir, ATLAS:2012qe}.
Therefore we will not discuss the searches for the vectorlike quarks anymore in the following.

\subsection{Neutralino NLSP}
\label{sec:chi0NLSP}
For the region with a neutralino NLSP, we interpret the ATLAS result of the search for the superparticles in events with no lepton, 2-6 jets and missing energy, with data from $\sqrt s = 8\TeV$ of $\int\!\mathcal{L} = 5.8\invfb$~\cite{ATLAS2012109}, as a constraint for the V-GMSB model.
The CMS collaboration provides a similar bound~\cite{recentcms}, which is expected to result in similar constraints on the model.

We employ Monte Carlo simulation to obtain constraints on the V-GMSB model.
The kinetic distribution is calculated by {\tt MadGraph 5}~\cite{MadGraph5} together with {\tt Pythia 6.4}~\cite{Pythia6.4} for simulating parton shower and initial/final state radiation, and {\tt Delphes 2.0}~\cite{Delphes} for detector simulation.
The events in $pp\to \tilde g\tilde g, \tilde g\tilde q^{(*)},\tilde q^{(*)}\tilde q^{(*)}$ channels, which are relevant for the SUSY signals, are generated~\footnote{%
The production channels with $\tilde t$ and $\tilde b$ are neglected, whose production cross sections are less than a few \% of the total cross section.}.

As for the object reconstruction, we follow the ATLAS procedure~\cite{ATLAS2012109} up to the {\tt Delphes} parameters.
The {\tt Delphes} original parameters for the ATLAS detector was used for the energy resolution and the calorimeter tower configuration.
Jets are reconstructed using the anti-$k_t$ algorithm~\cite{anti-kt,FASTJET} with $\Delta R = 0.4$.
Tracking efficiency for leptons ($e$ and $\mu$) is set 100\%, which results in tighter lepton veto compared to the original analysis and thus conservative limits~\footnote{%
Efficiency in lepton detection at the ATLAS experiment is estimated as $\sim90\%$ and $\sim95\%$ for electrons and muons, respectively, using $\sqrt s=7\TeV$ $pp$-collision data\cite{Aad:2011mk,ATLAS2011063}.
Including the lepton efficiency in the analysis results in an increase of the number of events which pass the following cuts; the increase is by $\sim 5\%$, which is smaller than theoretical uncertainties discussed later.
}.
Only the jets with $\pT>20\GeV$ and $|\eta|<2.8$, the electrons with $\pT>20\GeV$ and $|\eta|<2.47$, and the muons with $\pT>10\GeV$ and $|\eta|<2.4$ are considered.
For overlap removal, jets are rejected if electrons are found within a distance of $\Delta R = 0.2$, and then leptons within $\Delta R = 0.4$ of any jets are discarded.
Here, $\pT$ is a missing transverse momentum, 
$\Delta R \equiv \sqrt{(\Delta \eta)^2 + (\Delta \phi)^2}$ is a distance between two objects in the $(\eta,\phi)$ coordinate, and $\eta$ and $\phi$ are pseudo-rapidity and azimuthal angle around the beam direction, respectively.

Our analysis is employed with the same definition of the 12 inclusive signal regions (SRs) as the original paper~\cite{ATLAS2012109}, and we found that the ``D-tight'' SR among them yields the most stringent limit on the V-GMSB model (see Sec.~\ref{sec:LHC-result}). The selection procedure for the ``D-tight'' SR is defined as follows:
\begin{enumerate}
 \item having no lepton.
 \item having missing transverse energy $\MET>160\GeV$.
 \item having at least five jets with $\pT>60\GeV$. In addition, the leading jet have $\pT>130\GeV$.
 \item $\Delta\phi(\text{jet}, \MET)>0.4$ for all of the leading three jets and $>0.2$ for all jets with $\pT>40\GeV$.
 \item $\MET/m\s{eff}^{(5)} > 0.15$ and $m\s{eff}\suprm{inc} > 1700\GeV$.
\end{enumerate}
Here, the effective mass $m\s{eff}^{(5)}$ ($m\s{eff}\suprm{inc}$) is defined as the scalar sum of $\MET$ and $\pT$'s of the leading five jets (all jets with $\pT>40\GeV$).

The ATLAS search found no excess compared with the background estimation. The contribution from models beyond the SM (BSM) is constrained, which is shown as 95\% CL upper limit on the number of the events as $N\s{BSM}<6.0$.
This bound is interpreted as the constraint on the V-GMSB model.

In order to validate the above procedure, we have checked that the ATLAS result for the mSUGRA model in Ref.~\cite{ATLAS2012109} is reproduced in our simulation.
The calculated distribution of the effective mass coincides with  that of the ATLAS analysis.
The exclusion limit on the $m_0-m_{1/2}$ plane in our analysis is slightly looser than that of the ATLAS, but within $1\sigma$ range of the theoretical uncertainty.

\subsection{Stau NLSP}
Searches for heavy long-lived charged particles is relevant for the case with a quasi-stable stau NLSP.
We take the bound for the direct production of the lighter stau, $pp\to\tilde \tau_1\tilde \tau_1$, as the constraint on the V-GMSB scenario. The result is shown in the latest CMS report on the search for heavy long-lived charged particles with data at $\sqrt s=7\TeV$ and $\int\!\mathcal L=5.0\invfb$~\cite{Chatrchyan:2012sp}.
It gives a lower bound on the mass of the lighter stau as
\begin{equation}
m_{\tilde \tau_1}>223\GeV
\label{eq:stau223}
\end{equation}
at 95\% CL.
Utilizing {\tt Prospino}, we have checked that the cross section for the channel in the V-GMSB scenario is the same as that shown in the CMS report.

The CMS collaboration also published the constraints on typical GMSB models, where the charginos and/or neutralinos are 
the main production channels~\cite{Chatrchyan:2012sp}. However, it cannot be directly applied to the V-GMSB case, because the mass spectra are different between the two models (cf. Fig.~\ref{fig:spectrum}). The difference results in a deviation of the velocity distribution of the NLSP stau. Once the reconstruction efficiency of the heavy long-lived charged particles is published, the production channels of the charginos and neutralinos can provide a more strict constraint. Otherwise, the inclusion of these channels will suffer from large systematic uncertainties. On the other hand, the bound for the direct production of the lighter stau is independent of the model, and it is conservative. 
The ATLAS collaboration also published a result on searches for the same signals~\cite{:2012vd},
but their search assumes direct productions of the whole sleptons, i.e., the production is not limited to the lighter stau. The reconstruction efficiency of the particle is not provided either.
Since the mass spectrum of the V-GMSB model is different from the model analyzed in Ref.~\cite{:2012vd}, we will rely on the CMS bound on the direct stau production in Eq.~(\ref{eq:stau223}) in the following.


\subsection{Results}
\label{sec:LHC-result}

Numerical results for the LHC constraints on V-GMSB are summarized in Fig.~\ref{fig:results1}. 
Let us start from the neutralino NLSP region, i.e., below the light blue lines. 
The black solid lines represent the LHC constraint by searches for the jets with missing energy,
and the black dashed lines show the limit with theoretical uncertainties, for which we adopt $\pm 35\%$ errors in the production cross section. 
The uncertainties mainly come from errors of renormalization/factorization scales and a choice of the PDFs~\cite{Aad:2011ib}.
The left side of the black lines are excluded. Consequently, the gluino mass is required to be larger than about $1100\GeV$ for $M_{\rm mess}=10^6\GeV$, if the theoretical uncertainties are not included. The theoretical uncertainties can shift the mass bound by $\lesssim 100\GeV$.
When $M_{\rm mess}$ is larger, the exclusion becomes weaker for the fixed gluino mass. This is because the squarks become heavier, and the production cross section of the squark, especially that of $pp\to \tilde g\tilde q^{(*)}$, becomes smaller. When the messenger scale is as large as $10^{10}\GeV$, the bound becomes $m_{\tilde g} \gtrsim 1030\GeV$.

\begin{table}[t]
\begin{center}
\begin{tabular}{|c|c|c|c|c|c|c|}
\hline
$(\Lambda, M_{\rm mess}, \tan\beta)$
& \multicolumn{3}{|c|}{$(140\TeV, 10^6\GeV, 20)$}
& \multicolumn{3}{|c|}{$(150\TeV, 10^{10}\GeV, 20)$}
\\\hline
$m_{\tilde{g}}$
& 
\multicolumn{3}{|c|}{1116 GeV}
& 
\multicolumn{3}{|c|}{1146 GeV}
\\\hline
$m_{\tilde{q}_{\rm R}}$
& 
\multicolumn{3}{|c|}{1813 GeV}
& 
\multicolumn{3}{|c|}{2151 GeV}
\\\hline
production channel 
& $\tilde{g}\tilde{g}$& $\tilde{g}\tilde{q}$ & $\tilde{q}\tilde{q}$ 
& $\tilde{g}\tilde{g}$& $\tilde{g}\tilde{q}$ & $\tilde{q}\tilde{q}$ 
\\ \hline
cross section $\sigma$ (fb) 
& 5.54  & 2.75  & 0.238 
& 4.49  & 0.702 & 0.015
\\ \hline
acceptance $A$ 
& 0.0720 & 0.174 & 0.189 
& 0.0858 & 0.161 & 0.138
\\ \hline
$N=\sigma\times A\times 5.8{\rm fb}^{-1}$
& 2.31  & 2.77  & 0.261 
& 2.23  & 0.655  & 0.0120
\\\hline
$N$ (total)
& 
\multicolumn{3}{|c|}{5.3}
& 
\multicolumn{3}{|c|}{2.9}
\\ \hline
\end{tabular}
\end{center}
\caption{The cross section, the acceptance, and the expected number of events shown for the V-GMSB model points,
$(\Lambda, M_{\rm mess}, \tan\beta,M_V)=(140\TeV, 10^6\GeV, 20,1\TeV)$
and 
$(150\TeV, 10^{10}\GeV, 20,1\TeV)$.
The gluino mass, $m_{\tilde{g}}$, and the right-hand squark mass of the first generation, $m_{\tilde{q}_{\rm R}}$,
are also displayed. See Fig.~\ref{fig:spectrum} for the full mass spectrum of the former point. 
The acceptances are calculated for the ``D-tight'' SR, where 5000 events are generated at each model point.
}
\label{tab:gg_gq_qq}
\end{table}

The main production channels of SUSY events are $pp\to \tilde g\tilde g, \tilde g\tilde q,\tilde q\tilde q$. 
Although the squarks are relatively heavy compared to the gluino as shown in Fig.~\ref{fig:spectrum} and Fig.~\ref{fig:results1} (d), 
they are not decoupled from the productions.
Since the squark masses are almost independent of $\tan\beta$,
so are the LHC exclusion lines in Figs.~\ref{fig:results1} (a)-(c).
For illustration, in Table~\ref{tab:gg_gq_qq} we show the cross section of each channel at two model points,
$(\Lambda, M_{\rm mess}, \tan\beta,M_V)=(140\TeV, 10^6\GeV, 20,1\TeV)$
and 
$(150\TeV, 10^{10}\GeV, 20,1\TeV)$.
As can be seen in Fig.~\ref{fig:results1} (a) and (c), they are close to the LHC exclusion limit.
The acceptance and the expected number of events are also given in the ``D-tight'' SR. 
The acceptance is defined by a ratio of number of events that pass the cuts to the generated number of events.
It is found that the channels of $pp\to \tilde g\tilde g$ and $\tilde g\tilde q$
comparably contribute to the SUSY searches
for a low messenger scale (i.e., for a relatively light squark),
while $pp\to \tilde g\tilde g$ dominates for a high messenger scale (heavy squark).
However, the contribution of the $pp\to \tilde{q}\tilde{q}$ channel is small in the vicinity of the exclusion limits in Fig.~\ref{fig:results1} (a)-(c).

The ``D-tight'' cut requires five hard jets and large missing energy. It is easy to understand the origin of the four hard jets: the gluino is lightest among the colored superparticles, and it generates at least two hard jets when it decays into lighter superparticles.
The origin of the fifth jet can be understood as follows.
In the channel of $pp\to \tilde g\tilde q$, a hard fifth jet (or more jets) is generated by decays of the squarks into the gluino. In the $pp\to \tilde g\tilde g$ channel, 
the largest fraction of events which pass the ``D-tight'' cut have additional hard jet(s) from initial/final state radiations (ISR/FSR) (c.f.~\cite{Alwall:2009zu})~\footnote{
The hardness of the ISR and FSR jets might not be simulated correctly by {\tt Pythia}.
In order to resolve this uncertainty, we analyzed ``D-tight'' by means of the MLM-matching scheme~\cite{Mangano:2006rw} implemented into {\tt MadGraph} with a shower $k_t$ clustering and avoiding double counting between the gluino and squarks~\cite{Alwall:2008qv}. We have checked that the {\tt Pythia} calculation agrees with the MLM-matching result. 
}.
The ISR can yield additional hard jet(s) in the $pp\to \tilde g\tilde q$ channel as well.
In the rest of the events, the fifth jet comes from decays of $W^\pm$ and $\tau$ which are generated by the top quark decay and $\chi^{\pm}_1(\chi^0_2)\to \tau \tilde \tau \to \tau \tau \chi^0_1$ from the colored superparticle decay, respectively.

We have also checked that, in the mSUGRA model analyzed in Ref.~\cite{ATLAS2012109}, the parameter region which has squark and gluino masses similar to the V-GMSB model is mainly excluded by the ``D-tight'' SR. 
This result may support our conclusion that the V-GMSB model with a neutralino NLSP is mainly constrained by the ``D-tight'' SR among the 12 SRs of Ref.~\cite{ATLAS2012109}.

\begin{figure}[t!]
\begin{center}
\begin{tabular}{cc}
\includegraphics[width=0.46\linewidth]{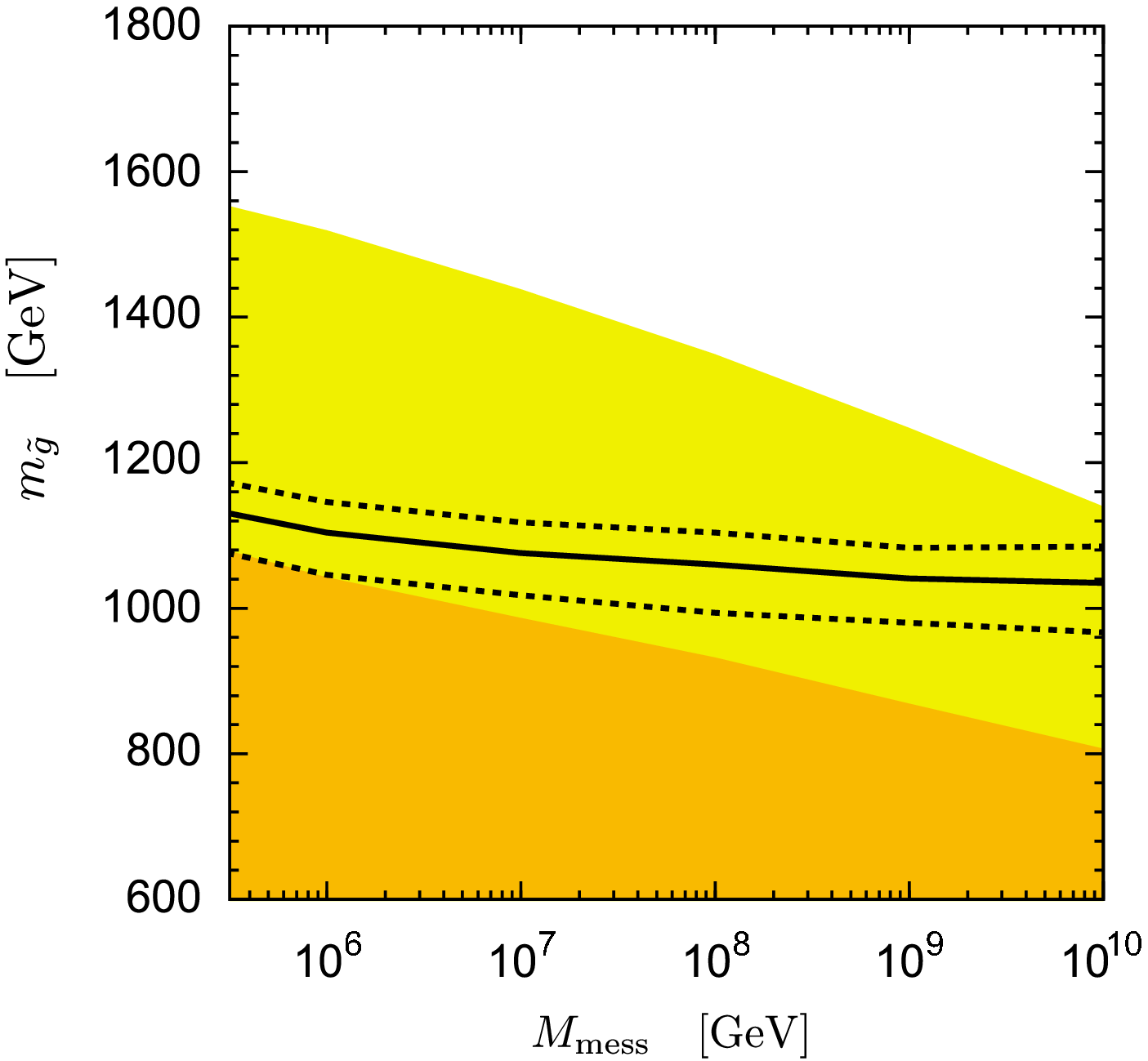}  &
\includegraphics[width=0.46\textwidth]{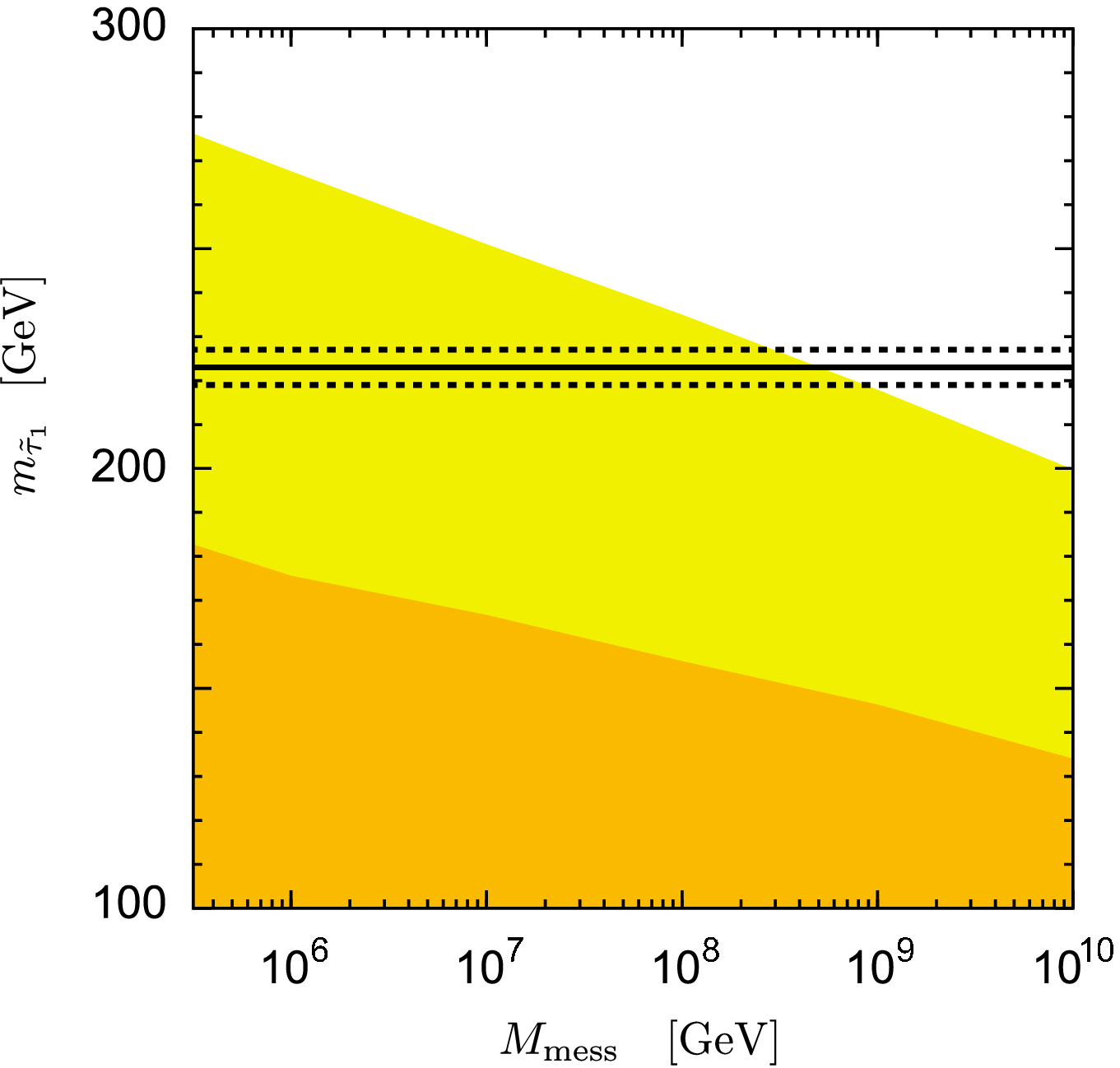}  
\end{tabular}
\end{center}
\caption{Lower bound from the LHC on the gluino mass and the stau mass when the NLSP is the neutralino (left) and the stau (right), respectively. 
In the yellow (orange) region, the muon $g-2$ discrepancy can be explained within $2\sigma$ ($1\sigma$).}
\label{fig:results2}
\end{figure}

In the stau NLSP region, i.e., above the light blue lines
in Fig.~\ref{fig:results1}, the black solid line indicates $m_{\tilde \tau_1}=223\,{\rm GeV}$, corresponding to the 95\% exclusion by searches for the heavy long-lived staus via the direct production~\cite{Chatrchyan:2012sp}.
The CMS collaboration evaluated the theoretical uncertainties of $3-7\%$ in the cross section calculations due to renormalization/factorization scales, $\alpha_s$ and PDFs. In the figure, we assigned theoretical error of 8\% for the cross section,
which corresponds to 2\% error for the stau mass bound, shown by black dashed lines. The left-side regions of the black lines are excluded.
Consequently, the region of the gluino mass lighter than $1200\GeV$ is excluded, and the bound becomes more severe for larger $\tan\beta$.
Since the theoretical uncertainty is small, this bound does not change so much even if it is taken into account.

It is interesting to analyze the LHC constraint as a function of the messenger scale in each category of the NLSP.
The results are shown in Fig.~\ref{fig:results2}.
The left and right panels correspond to the cases of the neutralino NLSP and the stau NLSP, respectively.
For a given $M_{\rm mess}$, the muon $g-2$ discrepancy can be explained at the $1\sigma$ ($2\sigma$) level when the gluino/stau mass is within the orange (yellow) region.
The upper ends of these regions represent the upper bound on the gluino/stau mass
in order to explain the muon $g-2$ at the $1\sigma$ ($2\sigma$) level.
In the case of the neutralino NLSP (left panel), the upper bound is determined by the requirement that the neutralino is lighter than the stau.
In fact, for a fixed value of the muon $g-2$, the gluino mass is maximized when $\tan\beta$ is set on the light blue lines in Fig.~\ref{fig:results1}.
When $M_{\rm mess}$ is as large as $10^{10}\GeV$, the vacuum stability bound can give a more severe bound. However, the result is almost unchanged, as noticed from Fig.~\ref{fig:results1} (a). 
This upper bound should be compared with 
lower bound on the gluino mass
from 
the LHC constraint, drawn by the black solid line.
Here, the exclusion limit is taken at $\tan\beta = 20$ as a reference, since the bound does not depend much on $\tan\beta$.
The black dashed lines include the theoretical uncertainties. (See discussion above.)
It is found that all the regions where the muon $g-2$ discrepancy is explained at the 1$\sigma$ level are already excluded by the direct searches for the superparticles at LHC.
The region that can explain the muon $g-2$ at 2$\sigma$ is still viable, 
and it is expected to be covered by LHC at $\sqrt{s}=13-14\TeV$.

For the stau NLSP case, the upper bounds on the stau mass are shown in the right panel of Fig.~\ref{fig:results2}. 
The upper ends of the yellow and orange regions are obtained by the requirement that the stau is the NLSP.
According to Fig.~\ref{fig:results1} (c) and (d), for a fixed muon $g-2$ the stau mass is maximized when it is close to the neutralino mass.
On the other hand, the black solid line is the lower bound on the stau mass from the LHC. 
It does not include the theoretical uncertainties; they are taken into account by the black dashed lines. 
Although the vacuum stability condition can give a tighter bound for $M_{\rm mess} \sim 10^{9-10}\GeV$, such a parameter region is already constrained by the LHC (see Fig.~\ref{fig:results1} (a)). 
Consequently, it is found that all the parameter region that can explain the muon $g-2$ at $1\sigma$ is excluded by the searches for the heavy long-lived charged particles at LHC, and it is expected to become more severe, for instance by analyzing the data at $\sqrt s=8\TeV$.

\section{Conclusion and Discussion}
\label{sec:discussion}

The discovery of the Higgs-like boson had a significant impact on the scenarios beyond the Standard Model, particularly on the SUSY models.
If the muon $g-2$ is a real signal of the SUSY,
the V-GMSB model is one of the only few viable SUSY models after the discovery of the 126 GeV Higgs-like boson.

In this paper, the collider searches for the superparticles in the V-GMSB model are explored in light of the current LHC results. The V-GMSB model predicts upper bounds on the gluino mass, i.e., the superparticle mass scale, to explain the muon $g-2$ discrepancy with realizing the Higgs boson mass of $126\GeV$. The LHC signatures are characterized by the NLSP. In the V-GMSB, the candidates are the neutralino and the stau. 
It was shown that the parameter regions of the gluino mass of $<1000-1100\GeV$ are excluded in the case when the neutralino is the NLSP, 
and the constraint is more severe for the stau NLSP. 
In both cases, it was shown that the whole parameter regions where the muon $g-2$ is saturated at the 1$\sigma$ level are already excluded by the searches for the jets with missing transverse energy or those for the heavy long-lived charged particles. When the neutralino is the NLSP, all the region favored by the muon $g-2$ at 2$\sigma$ will be covered by LHC after the upgrade for $\sqrt{s}=13-14\TeV$, while the whole region of the muon $g-2$ at 2$\sigma$ is expected to be covered earlier for the stau NLSP case. 

As shown in this paper, it is rather difficult to explain the muon $g-2$ at 1$\sigma$ for the long-lived NLSP cases. One of the remaining possibilities is that the stau is the NLSP with a decay length less than $\mathcal{O}({\rm cm})$. This short decay length is realized when the R-parity is slightly violated or the gravitino is as light as $\mathcal{O}(10-1000)$\,eV. GUT breaking effects on the SUSY breaking (invariant) masses of the doublet and triplet messengers can also help to avoid the constraints from the LHC results and to enhance the SUSY contributions to the muon $g-2$ by generating more hierarchical masses of the colored and non-colored SUSY particles.

Let us finally comment on the cosmology of the V-GMSB model. 
Assuming R-parity conservation, the gravitino LSP is completely stable, and the NLSP becomes long-lived  unless the gravitino is ultra-light. The late-time decay of the NLSP may spoil the success of the big-bang nucleosynthesis (BBN), 
and hence there are constraints on the lifetime and the abundance of the NLSP, 
which translate into upper bounds on the gravitino mass, depending on the species and the mass of the NLSP.
In Fig.~\ref{fig:results1},  the NLSP mass is bounded from above in the region where the discrepancy in the muon $g-2$ is reduced to $2\sigma$ level; $m_{\rm NLSP} < 270$ GeV for both of the neutralino and stau NLSP cases,   
for $M_{\rm mess}>10^6\GeV$. 
The BBN constraints on the gravitino mass are correspondingly $m_{3/2} \lsim 0.4 \GeV$ for the neutralino NLSP and $m_{3/2} \lsim 4\GeV$ for the stau NLSP~\cite{Kawasaki:2008qe}. 
The gravitino LSP with a mass of $\mathcal{O}(1)\GeV$ can be the dominant component of the dark matter
if the reheating temperature is as high as $T_R = 10^6-10^7$ GeV~\cite{gravitino_and_reheating}~\footnote{The BBN constraints and the gravitino abundance depend on the mass spectrum of the MSSM and vectorlike particles.
In particular, since the SU(3) gauge coupling remains strong up to the high energy scale in V-GMSB model, the perturbation in the calculation of the gravitino production becomes troublesome~\cite{gravitino-strong}, which requires an improved analysis to reliably calculate the gravitino abundance.}.
Then, the non-thermal leptogenesis~\cite{nonth-LG} may marginally work.
We should also note that the notorious inflaton-induced gravitino problem~\cite{Endo:2007sz}, which excludes many inflation models in the gravity-mediated SUSY breaking scenario with $\mathcal{O}(1)$ TeV gravitino, can be avoided in (V-)GMSB models.
Furthermore,  the Polonyi problem can also be avoided in the GMSB models~\footnote{Recently, a low energy SUSY model which can explain the Higgs boson mass of around 126 GeV without the Polonyi problem and flavor problem was proposed in Ref.~\cite{Moroi:2012kg}.};
rather, the decay of the coherent oscillation of the SUSY breaking field (Polonyi field) may lead to an alternative viable cosmological scenario~\cite{nonthermalGravitinoDMinGMSB}.
More detailed study of cosmological scenarios in the V-GMSB model is beyond the scope of the present work, and will be given elsewhere.

\section*{Acknowledgment}
We thank Kohsaku Tobioka, Keisuke Harigaya, Shigeki Matsumoto, Mihoko M. Nojiri, Kazuki Sakurai, Teppei Kitahara for useful discussions.
This work was supported by JSPS KAKENHI Grant 
No.~23740172 (M.E.), No.~21740164 (K.H.), No.~22244021 (K.H.), No.~22-8132 (S.I.) and No.~24-7523 (N.Y.). 
This work was supported by World Premier International Research Center Initiative (WPI Initiative), MEXT, Japan.


{\small

\begingroup\raggedright\endgroup
}
\end{document}